\documentclass[runningheads]{llncs}
\usepackage[T1]{fontenc}
\usepackage{graphicx,verbatim}
\usepackage{amsmath}
\usepackage{booktabs}
\usepackage{xcolor}

\begin{document}

\title{A Lightweight Optimization Framework for Estimating 3D Brain Tumor Infiltration}

\titlerunning{Estimating 3D Brain Tumor Infiltration}
\institute{Princeton University, Princeton NJ 08544, USA \and
Springer Heidelberg, Tiergartenstr. 17, 69121 Heidelberg, Germany
\email{lncs@springer.com}\\
\url{http://www.springer.com/gp/computer-science/lncs} \and
ABC Institute, Rupert-Karls-University Heidelberg, Heidelberg, Germany\\
\email{\{abc,lncs\}@uni-heidelberg.de}}

\author{Jonas Weidner\inst{1,2} \and Michal Balcerak\inst{3} \and Ivan Ezhov\inst{1, 2} \and André Datchev\inst{1} \and Laurin Lux\inst{1,2} \and Lucas Zimmer\inst{1,2} \and Daniel Rueckert\inst{1,2,4} \and Björn Menze\inst{3} \and  Benedikt Wiestler\inst{1,2}}
\authorrunning{Weidner et al.}

\institute{Technical University of Munich\\ \and Munich Center for Machine Learning \\ \and University of Zurich\\ \and Imperial College London \\ \email{j.weidner@tum.de}}

\maketitle              
\begin{abstract}

Glioblastoma, the most aggressive primary brain tumor, poses a severe clinical challenge due to its diffuse microscopic infiltration, which remains largely undetected on standard MRI. As a result, current radiotherapy planning employs a uniform 15 mm margin around the resection cavity, failing to capture patient-specific tumor spread. Tumor growth modeling offers a promising approach to reveal this hidden infiltration.
However, methods based on partial differential equations or physics-informed neural networks tend to be computationally intensive or overly constrained, limiting their clinical adaptability to individual patients. 
In this work, we propose a lightweight, rapid, and robust optimization framework\footnote{"github.com/jonasw247/spatial-brain-tumor-concentration-estimation"} that estimates the 3D tumor concentration by fitting it to MRI tumor segmentations while enforcing a smooth concentration landscape.
This approach achieves superior tumor recurrence prediction on 192 brain tumor patients across two public datasets, outperforming state-of-the-art baselines while reducing runtime from 30 minutes to less than one minute.
Furthermore, we demonstrate the framework's versatility and adaptability by showing its ability to seamlessly integrate additional imaging modalities or physical constraints.

\end{abstract}

\section{Introduction}
The treatment of glioblastoma, the most aggressive primary brain tumor, presents a severe clinical challenge with persistently low survival rates. A key reason for this is the diffuse infiltration of tumor cells into the surrounding brain, making them the primary targets of postoperative radiotherapy. However, this infiltration remains largely undetected in standard magnetic resonance imaging (MRI), while current radiotherapy treatment planning relies on a simple protocol \cite{niyazi2023estro}: A uniform 15\,mm margin is drawn around the resection cavity to account for microscopic tumor cell infiltration.
\begin{figure}[ht!]
    \centering
    \includegraphics[width=1\textwidth]{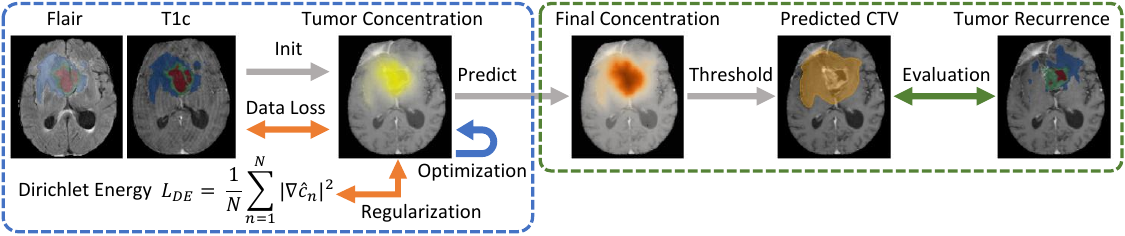}
    \caption{We optimize (blue) a 3D scalar tumor concentration estimation (yellow) by simultaneously fitting the data while ensuring a smooth concentration landscape by minimizing the Dirichlet energy. Using this predicted tumor concentration (orange), we propose a radiotherapy plan (Clinical Target Volume (CTV), orange). We evaluate (green) our method's ability to capture areas of subsequent tumor recurrence.}
    \label{fig:overview}
\end{figure}

Tumor growth modeling promises to reveal the otherwise invisible tumor cell infiltration. Traditionally, pure biophysical models were calibrated against real imaging data with compute-intensive sampling approaches \cite{lipkova2019personalized}. Learning-based methods \cite{ezhov2023learn,zhang2025personalized} have shown dramatic speedups but often lack robustness and precision due to a limited amount of data. 
Combining both approaches by using a deep learning  prior and subsequent sampling reduces runtime and enhances precision but still falls short of the runtime achieved by purely deep learning-based methods \cite{Weidner2024}. Alternatively, path-based methods like \cite{bortfeld2022modeling} provide a simple geometric evolution but ignore the physical properties of tumor progression. Recently, \cite{balcerakphysics,balcerak2023individualizing} demonstrated that advanced models, incorporating brain deformation, several imaging modalities, and physical tumor properties, can precisely capture the tumor growth process while outperforming conventional methods in tumor recurrence prediction. These methods fit a discrete 4D (3D plus time) tumor cell concentration to the data while preserving a physically plausible growth process. They are considered the best-performing approach demonstrated by tumor recurrence prediction. However, the complex models require extensive heuristic fine-tuning and long runtimes, hindering clinical approval and adaptation.


Here, we propose an efficient and flexible optimization framework to estimate the hidden 3D tumor concentration by fitting it to the MRI tumor segmentation while simultaneously ensuring a smooth concentration landscape, providing \textbf{the following contributions}:

\begin{enumerate}

\item  We introduce a lightweight, rapid, and robust optimization method for planning brain tumor radiotherapy target volumes, which \emph{outperforms the current clinical standard of care.}
\item  We demonstrate that our optimization method \emph{improves recurrence prediction} over competing state-of-the-art methods.
\item  We showcase our method's \emph{extensibility and adaptability} by including different imaging modalities like PET and additional physical constraints.
\end{enumerate}

\section{Methods}
We propose an optimization of tumor cells spread in the brain (Figure \ref{fig:overview}). 
We aim to fit the tumor cell concentration for each voxel to the data while simultaneously following a physically plausible distribution. Therefore, we optimize the concentration based on the gradients of differentiable loss terms. 

\subsubsection{Loss:}
Our framework combines two distinct losses $L$ with weights $\lambda$. While the \emph{data loss} $L_{\text{Data}}$ closely matches the predicted tumor cell concentration to the visible tumor in the MRI, the \emph{Dirichlet Energy loss} $L_{\text{DE}}$ ensures that the proposed final tumor concentration follows a physically plausible, continuous distribution: \begin{equation}
L_{\text{Total}} = \lambda_\text{Data} L_{\text{Data}} + \lambda_\text{DE} L_{\text{DE}}
\end{equation}

MR images are used to calculate data loss. We use contrast-enhanced T1c and Flair images to segment the tumor in enhancing tumor and necrotic areas (combined into $S_{\text{Core}}$), and edema ($S_{\text{Edema}}$) using BraTS.Toolkit \cite{kofler2020brats}. To assess the data loss, we apply the Dice score ($DSC$) \cite{dice1945measures} between the predicted tumor concentration above a visibility threshold and the segmentation derived from the MRI: $L_{\text{Data}} =  \, \alpha_{\text{Core}} \, DSC(\hat{c} > \tau_{\text{Core}},\ S_{\text{Core}})  \nonumber  + \alpha_{\text{Edema}} \, DSC(\hat{c} > \tau_{\text{Edema}},\ S_{\text{Edema}} \cup S_{\text{Core} }) $.
We set thresholds for the estimated tumor concentration $\hat{c}$ based on predefined values for the tumor core ($\tau_{\text{Core}}$) and edema ($\tau_{\text{Edema}}$) \cite{lipkova2019personalized}, below which we assume the tumor is not detectable in the MRI.

The growth of a tumor is typically modeled by a reaction-diffusion partial differential equation (PDE), like the Fisher-Kolmogorov equation: $\frac{\partial c}{\partial t} = \nabla \cdot (D \nabla c) + \rho c (1 - c)$.
The temporal development of the tumor concentration $c$ is described by the logistic growth parameter $\rho$ and the diffusion coefficient $D$. In literature, many variations of this equation exist using different growth terms, different tumor diffusion coefficients for other tissue types, and additional terms like advection \cite{subramanian2019simulation,saut2014multilayer}.

Reaction-diffusion equations yield well-behaved solutions in space and time, characterized by continuous changes and the absence of abrupt variations. Inspired by this, we penalize the Dirichlet energy of the tumor concentration field. It is defined as the $L_2$ norm of the gradients over all voxels $N$: $ L_{\text{DE}} = \frac{1}{N} \sum_{n = 1}^{N} |\nabla \hat{c}_n|^2$.

\subsubsection{Extending our framework:}
Besides its primary contribution, the combined data \& Dirichlet Energy loss function, our framework is designed to be expandable through additional loss terms that add specific objectives or constraints. Here, we demonstrate how advanced imaging modalities can be included as additional data loss and how physical constraints may be utilized.

\underline{Including PET Imaging:}
Besides the most common MR sequences, Flair and T1c, from which we extract the edema and core segmentations, several other imaging modalities are typically acquired. In this example, we focus on positron emission tomography (PET) imaging. It was shown that PET image intensity $I_\text{PET}$ correlates with tumor concentration inside the edema region \cite{liesche2021visualizing}. To incorporate this into our estimation of tumor cell concentration, we define the additional loss term as: $L_{\text{PET}} = \text{corr}(\hat{c}, I_\text{PET} )$.

\underline{Including Physical Constraints:} 
It can be shown that a traveling wave in the form of a sigmoid solves the Fisher-Kolmogorov equation in one dimension \cite{swanson2008mathematical}.  

\begin{equation}
    c(x, t) = \left(1 + \exp\left(\frac{x - 2 \sqrt{D \rho }t}{\sqrt{\frac{D}{\rho}}}\right)\right)^{-1} \Rightarrow   \frac{\partial c}{\partial x} = \sqrt{\frac{\rho}{D}} c (1-c)
    \label{eq:Analytical_solution_dif}
\end{equation}

After a tumor grows to a large size, compared to the origin cell concentration, the exact initial condition becomes unimportant \cite{konukouglu2006extrapolating}.
We assume this sigmoid behavior locally at a given time. 
Thus, we can conclude that the following condition must be fulfilled for the predicted tumor concentration: 
$|\nabla  \hat{c}| \stackrel{!}{=} \left|\frac{\partial c}{\partial x}\right|$. 
If we plug in the analytical solution  (equation \ref{eq:Analytical_solution_dif}), $    |\nabla  \hat{c}| = \left|  \sqrt{\frac{\rho}{D}} \hat{c} (1-\hat{c}) \right|
    \label{eq:physCond}$ follows.
For isotropic, homogeneous brain tissue, the parameters  $\rho$ and $D$ can be approximated by the constant $ k\approx \sqrt{\rho / D} $ representing the tumor’s slope fading. The loss is defined as the mean squared error of the residuals, normalized by k: $   L_{\text{Wave}} =\frac{1}{N} \sum_{n = 1}^{N} ( \frac{1}{k} (|\nabla \hat{c}_n| - k \hat{c}_n (1-\hat{c}_n) ))^2$.
The normalization is done to prevent the simple solution of a homogeneous tumor concentration.

\subsubsection{Optimization:}
The tumor concentration in each voxel is optimized after initializing it based on the thresholds of the tumor core and edema, ensuring minimal data loss. The initial concentrations are set as $c_{i_\text{Core}} = \tau_\text{Core} + 0.01$, $c_{i_\text{Edema}} = \tau_\text{Edema} + 0.01$, $c_{i_\text{Brain}} = 0.01$ for the rest of the brain. The slope for the wave loss was initialized as $k_{i_\text{Wave}} = 0.1 \text{ mm}^{-1}$.
The loss weights were empirically found to be optimal when the loss terms were of a similar order of magnitude: $\alpha_\text{Core} = 1$, $\alpha_\text{Edema} = 1$, $\lambda_\text{Data} = 1$, $\lambda_\text{DE} = 1000$, $\lambda_\text{Wave} = 1000$, and $\lambda_\text{PET} = 1$. 

 We sampled the thresholds $\tau_\text{Core}$ from $\{0.6,\allowbreak 0.7, \allowbreak 0.8, \allowbreak 0.9\}$ and $\tau_\text{Edema}$ from $\{0.1, \allowbreak 0.2, \allowbreak 0.3, \allowbreak 0.4, \allowbreak 0.5\}$. The optimization process runs for $500$ steps. 
 As it is hard to define clear thresholds $\tau_{\text{Core}}$ and $\tau_{\text{Edema}}$ for the visible tumor concentration, we decided to test a wide range of clinically plausible choices. 
 We utilize the highly optimized PyTorch GPU implementation of the Adam optimizer. 
 
\subsubsection{Evaluation:}
To evaluate the estimated tumor concentration, we test its ability to predict tumor recurrence. We assume that post-operative tumor recurrence is correlated with preoperative tumor cell concentration. Importantly, this metric also directly links our method with informing individualized radiotherapy plans and thus measures its clinical utility.

For our experiments, we require both the preoperative MRI (to estimate tumor cell concentrations) and a later follow-up MRI showing tumor recurrence, registered to the preoperative space. We use the following two public datasets: \textbf{GliODIL} contains 152 glioblastoma patients \cite{balcerak2023individualizing}, with estimated tissue concentrations and segmentations of tumor and tumor recurrence. FET-PET imaging is available for a subset of 58 patients. \textbf{RHUH} contains pre-operative, post-operative, and recurrence MR images of 40 patients and segmentations \cite{cepeda2023rio}. We employed ANTs with the optimized settings from the BraTSReg challenge \cite{baheti2021brain} to deformably register recurrence into preop space.

We create synthetic standard plans following the current clinical guidelines \cite{niyazi2023estro} in the same way as done by \cite{balcerak2023individualizing}. We use the volume of the tumor core as an approximation for the resection cavity and add an additional, isotropic 15 mm margin around it. We construct a proposed CTV with the \textit{same volume} as this standard plan for each compared method. Then, we measure the percentage of tumor recurrence covered by this binarized volume, distinguishing between the prediction of "enhancing recurrence" and "any recurrence", which is the area of edema, necrotic, or enhancing recurrence.  

We always compare paired data for different methods. The results are typically not normally distributed, as the recurrence is often covered completely by all methods or is not covered at all. Thus, we used the paired Wilcoxon-ranked test. We marked significant differences to the standard plan with "$*$" for $p<0.05$ and "$**$" for $p< 0.01$. For comparison to the next best baseline method, "static grid discretization", we use "$\dagger$" and "$\ddagger$". 

We compare our method to the following: 
\textbf{Standard Plan} refers to the standard procedure defining the recurrence prediction volume for all other methods. A 15 mm margin is constructed around the enhancing recurrence. \textbf{Numerical Physics Simulations}, which utilizes the evolutionary sampling strategy of numerically simulated tumors \cite{Weidner2024}. \textbf{Data-Driven Neural Networks (Unconstrained)} describes the UNet adaptation introduced by \cite{balcerak2023individualizing}. \textbf{Data-Driven Neural Networks (Physics-Constrained)} refers to the neural inverse solver by \cite{ezhov2023learn}. \textbf{Static Grid Discretization} optimizes the 3D + time tumor growth process on a static grid constrained by physically plausible timesteps \cite{balcerak2023individualizing}. \textbf{Deformable Grid Discretization} extends the Static Grid Discretization by a deformable mesh and PET image information \cite{balcerakphysics}.


\section{Results} 
\begin{figure}[ht!]
    \centering
    \includegraphics[width=\textwidth]{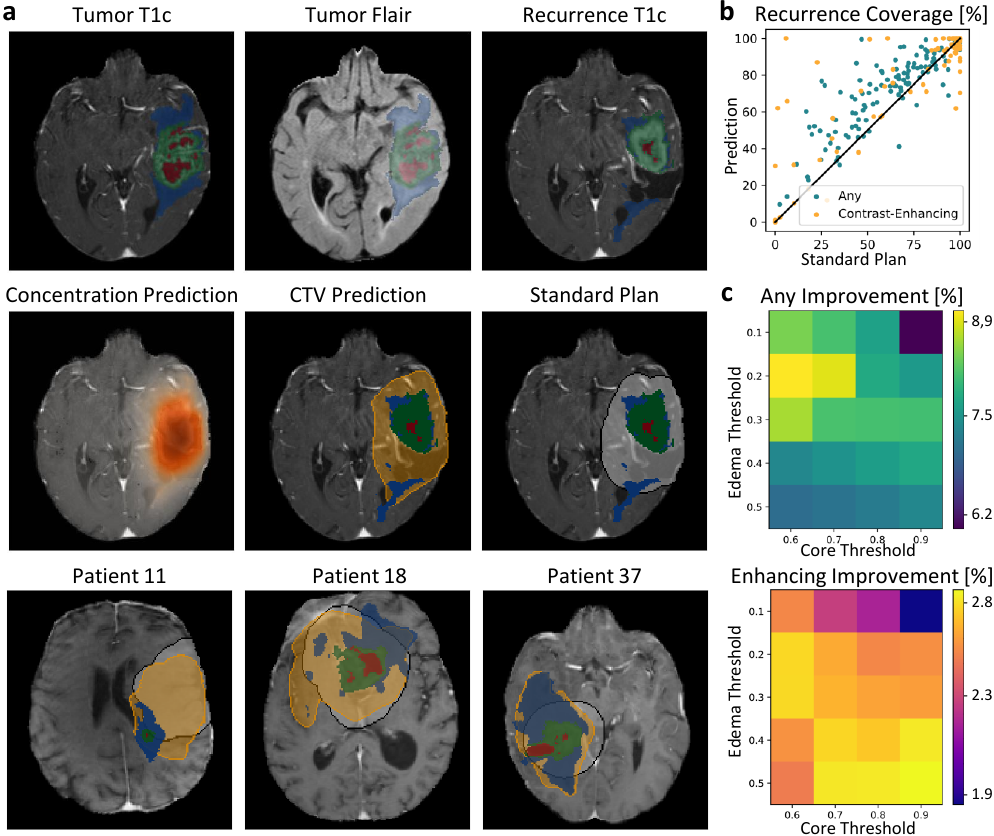}
    \caption{a.) Demonstration of our method on example patients. In the first row, we show the two input MR images with the tumor and the recurrence that should be covered. Edema is shown in blue, enhancing tumor in green, and necrotic in red. Our method predicts a continuous estimation of tumor cells, as shown in the second row. This continuous concentration is thresholded to have the same volume as the standard plan (grey) to create the CTV (orange). In the last row, we compare our method to the standard plan for different RHUH patients. b.) The recurrence coverage is shown for the GliODIL Dataset with 152 patients. The individual results for each patient are shown compared to the standard plan. A clear improvement is visible for many patients, while also a lot of patients result in 0\% or 100\% coverage. c.) We are comparing any recurrence (green) and contrast-enhancing recurrence (orange) coverage improvement over the standard plan for different core and edema visibility threshold parameters $\tau$.}
    \label{fig:exampleImages}
\end{figure}

We evaluate our results mainly on the GliODIL dataset with 152 patients. 
A qualitative assessment of how our method compares against the current clinical gold standard is shown in Figure \ref{fig:exampleImages}. 
Although it is ignored in the clinical standard, there is a clear indication that edema has a predictive correlation with tumor recurrence. This is also the most significant benefit over the best baseline method, "static grid discretization".  

As it is unclear which tumor concentrations are visible in MR images, we tested an extensive, medically plausible range of thresholds $\tau_{\text{Core}}$ and $\tau_{\text{Edema}}$. For our comparison on recurrence prediction, these tumor thresholds can also be interpreted as the weighting between the importance of tumor core \textit{vs.} edema. We show the sweep over those thresholds in Figure \ref{fig:exampleImages}c. In general, we do not find a large variation for different thresholds. Predicting the enhancing recurrence is slightly easier when selecting a high core threshold. For the "any recurrence" prediction, we find an optimal outcome at ($\tau_\text{Core} = 0.6$, $\tau_\text{Edema} = 0.2$). 

The quantitative results are shown in Table \ref {table:validation_combined}, where we compare our data to the state-of-the-art models, \cite{balcerak2023individualizing,ezhov2023learn,Weidner2024}. 
We see that our method significantly outperforms all existing methods in predicting "any recurrence", meaning edema, necrosis, or enhancing core, and also in predicting only contrast-enhancing tumor recurrence while only partly significant. 

To validate our findings, we compare our method with the standard plan and the best baseline method on the independent RHUH dataset, which stems from another center. The results are shown in Table \ref{table:pet}a. We find similar results to those on the primary dataset. For "any recurrence", we see a significant improvement of 5\% to 8\% compared to the standard plan. Identical to the GliODIL dataset, the improvement in predicting the contrast-enhancing part of the results is about 2\% to 3\%.

\begin{table}[ht!]
  \caption{Comparison of recurrence segmentation coverage given equal radiation volume, tested for different edema and core thresholds (Figure \ref{fig:exampleImages}) on the GliODIL dataset with 152 patients. Our method outperforms all others with short runtime.}
  \label{table:validation_combined}
  \centering
  \scriptsize
  \resizebox{1\textwidth }{!}{%
  \begin{tabular}{lccc}
    \toprule
    \textbf{Recurrence Coverage - GliODIL}& \textbf{Any [\%]} & \textbf{Enhancing Core [\%]} & \textbf{Runtime} \\
    \midrule
    
    NN (Unconstrained) \cite{balcerak2023individualizing}&  $65.38 \pm 2.05$ &$69.02 \pm 2.79$ & < 1 min\\
    NN (Physics-Constrained) \cite{ezhov2023learn}& $ 62.06 \pm 2.11$ & $75.25 \pm 2.84$ & < 1 min \\
    Numerical Physics Simulations  \cite{Weidner2024}& $61.16 \pm 2.12$ & $75.34 \pm 2.87$ & 2 h\\
    
    Static Grid Discretization \cite{balcerak2023individualizing} &$67.80 \pm 2.09$  & $84.42 \pm 2.40$ & 30 min\\
    \midrule
    Standard Plan & $63.59 \pm 2.26$&$82.42\pm 2.60 $ & < 1 min\\
    \midrule
    Ours Worst Thresholds &  $ 69.72\pm 2.07^{\ddagger**}$ & $84.34 \pm 2.38^{**}$ &1 min\\
    Ours Median Thresholds &  $ 70.93 \pm 1.99^{\ddagger**}$ & $85.02 \pm  2.31^{}$ &1 min\\
    Ours Best Thresholds  & $ \mathbf{72.48} \pm 1.99^{\ddagger**}$ & $\mathbf{85.19 } \pm 2.28^{}$ & 1 min\\
    




    \bottomrule
  \end{tabular} 
  }
\end{table}



\subsubsection{Including PET Imaging:}
As an example of how our method can be extended with additional measurements, we evaluate the effects of an additional PET loss. 
We tested 58 patients with amino acid PET images in the GliODIL dataset. The same experiment was conducted by \cite{balcerakphysics} in a recent NeurIPS publication, simulating not only the influence of PET but also advanced brain deformations. In Table \ref{table:pet}, we compare the default method with optimal parameters and our method with the additional PET loss to this method and the other methods also used for the full GliODIL dataset (Table \ref{table:validation_combined}). Both the default method and the one including PET loss show a clear improvement over the standard plan and the dynamic grid discretization regarding "any recurrence" prediction. In enhancing recurrence coverage, the dynamic grid discretization is better than our method, while our methods are still outperforming the clinical standard plan. 

\begin{table}[ht!]
  \caption{a.) We evaluated our method on the independent second RHUH Dataset with 40 patients. We compared our method to the standard plan and the best-performing baseline from the GliODIL dataset. b.) Comparison of our additional method utilizing PET imaging. We use the subset of 58 patients of the Gliod dataset having PET imaging. Additionally, we can compare our results to the method "Dynamic Grid Discretization"\cite{balcerakphysics}, which requires PET. c.) Recurrence prediction with the additional assumption of a wave-like solution compared to the default version of our method. We compare the minimum, maximum, and median of recurrence coverage with multiple thresholds as shown in Figure \ref{fig:exampleImages}.}
  \label{table:pet}
  \centering
  \scriptsize
  \resizebox{1\textwidth }{!}{%
  \begin{tabular}{lcc}
    \toprule
    \textbf{Experiment - Recurrence Coverage}& \textbf{Any [\%]} & \textbf{Enhancing Core [\%]} \\
    \midrule
    \textbf{a.) RHUH (40 Patients) } \\
    Static Grid Discretization & $70.95 \pm 3.29$  & $79.59 \pm 4.82$\\
    Standard Plan &$65.47\pm 3.51 $& $80.69 \pm 5.07$\\ 

    Ours Worst &  $ 70.27\pm 3.28^{**}$ & $80.70 \pm 4.60$\\
    Ours Median &  $ 72.02 \pm 3.21^{**}$ & $ 81.89\pm 4.73$ \\
    Ours Best &$\mathbf{73.18} \pm 3.16^{\dagger**}$ \ \ & $\mathbf{82.06} \pm 4.70 $  \\




    \midrule
    \textbf{b.) With PET Images (58 Patients)}& & \\
    NN (Unconstrained) & $59.0 \pm 4.3$ & $66.8 \pm 4.9$ \\
    NN (Physics-Constrained) & $70.4 \pm 3.7$ & $84.3 \pm 3.3$ \\
    Numerical Physics Simulations  & $67.1 \pm 3.8$ & $86.2 \pm 3.6$ \\
    
    Static Grid Discretization& $72.9 \pm 3.5$ & $89.0 \pm 3.3$ \\
    Dynamic Grid Discretization & $74.7 \pm 3.1$ &$\mathbf{89.9} \pm 2.7$ \\
    Standard Plan & $70.0 \pm 3.8$ & $87.3 \pm 3.6$ \\
    Ours & $76.2 \pm 3.4^{\ddagger**}$ & $88.2 \pm 3.4$\\
    Ours with additional PET Loss &   $\mathbf{77.4}\pm 3.3^{\ddagger**}$  & $88.4 \pm 3.3$\\
    \midrule



    \textbf{c.) With Wave Loss (152 Patients)} \ \ \ \ \ \ && \\
    
    
    Ours Worst Thresholds &  $ 69.72\pm 2.07^{\ddagger**}$ & $84.34 \pm 2.38^{**}$ \\
    Ours Median Thresholds &  $ 70.93 \pm 1.99^{\ddagger**}$ & $85.02 \pm  2.31$ \\
    Ours Best Thresholds  & $ \mathbf{72.48} \pm 1.99^{\ddagger**}$ & $85.19  \pm 2.28$ \\

    Ours with Wave Loss Worst Thresholds &  $ 71.24\pm 2.08^{\ddagger**}$ & $84.79 \pm 2.34$ \\
    Ours with Wave Loss Median Thresholds &  $ 71.79 \pm 2.07^{\ddagger**}$ & $85.03 \pm  2.35^{\ddagger*}$ \\
    Ours with Wave Loss Thresholds  & $ 72.16\pm 2.06^{\ddagger**}$ & $\mathbf{85.46 } \pm 2.32^{\ddagger**}$ \\
    \bottomrule


  \end{tabular} 
  }
\end{table}

\subsubsection{Including Physical Constraints:}
We also test how the assumption of a wave-like solution, i.e., adding further physical constraints to the estimated tumor cell concentration, affects the recurrence prediction. We ran the same tumor visibility parameter sweep as shown in Figure \ref{fig:exampleImages} with the additional wave loss. The results are compared to the default loss and are shown in Table \ref{table:validation_combined}. We see a clear improvement for the median and the worst set of thresholds in predicting "any recurrence". Additionally, by adding the wave loss, we obtain a significant improvement in the prediction of enhancing core recurrence. This can be interpreted as increased robustness towards the selection of hyperparameters, paving the way for further adjustment of advanced physics constraints and highlighting how extending our framework with additional physical constraints can improve model safety, an important prerequisite for clinical translation.

\section{Conclusion}

Our method estimates the 3D concentration of tumor cells by aligning it with MRI tumor segmentation while enforcing a smooth concentration landscape. This approach cuts runtimes from 30 minutes to under one minute and significantly improves recurrence prediction for 192 patients. This streamlined approach challenges the demand for complex simulations by demonstrating that simpler, efficient techniques can effectively address the challenging problem of tumor recurrence prediction. 
Our experimental results indicate that additional loss terms, particularly the inclusion of extra physical constraints, can stabilize predictions and provide essential safety guarantees, which are critical for clinical translation and effective radiotherapy planning. Also, this highlights the flexibility of our framework, which can thus be readily adapted to other pathologies and modalities.


\newpage
\bibliographystyle{splncs04} 
\bibliography{bib.bib}

\end{document}